\documentclass{kluwer}
\usepackage{epsf}
\usepackage[dvips]{epsfig}

\begin{document}
\begin{article}

  \begin{opening}
    \title{Merging Timescales and Merger Rates of Star Clusters in
      Dense Star Cluster Complexes}
    \author{M. \surname{Fellhauer}}
    \institute{Astronomisches Rechen-Institut, Heidelberg,
      Germany} 
    \author{H. \surname{Baumgardt}}
    \institute{Department of Mathematics \& Statistics,
      University of Edinburgh, Scotland, UK}
    \author{P. \surname{Kroupa}}
    \institute{Institut f\"{u}r theor. Astrophysik, University of
      Kiel, Germany}
    \author{R. \surname{Spurzem}}
    \institute{Astronomisches Rechen-Institut, Heidelberg,
      Germany} 
    \date{\today}
    \runningtitle{Merging Star Clusters}
    \runningauthor{Fellhauer et al.}

    \begin{abstract}
      Interacting galaxies like the famous Antennae
      (NGC~4038/4039) or Stephan's Quintet (HCG~92) show
      considerable star forming activity in their tidal arms.
      High resolution images (e.g. from HST-observations)
      indicate that these regions consist of up to hundreds of
      massive stellar clusters or tidal dwarf galaxies (TDG).  In
      this paper we want to investigate the future fate of these
      clusters of massive star clusters (in this work called {\em
      super-clusters}).  We simulate compact super-clusters in
      the tidal field of a host-galaxy and investigate the
      influence of orbital and internal parameters on the rate
      and timescale of the merging process.  We show that it is
      possible that such configurations merge and build a dwarf
      galaxy, which could be an important mechanism of how
      long-lived dwarf satellite galaxies form.  A detailed study
      of the merger object will appear in a follow-up paper.
    \end{abstract}
  \end{opening}

\section{Introduction}
\label{sec:intro}

High resolution images from HST-observations of the Antennae
galaxies (Whitmore et al.\ 1999, Zhang \& Fall 1999) show that
the star forming regions there consist of up to hundreds of young
(ages 3--7~Myr), compact massive star clusters with dimensions of
a few pc.  The young star clusters in the Antennae have effective
radii $R_{\rm eff} = 4$~pc and masses of the order $10^{4}$ --
$10^{6}$~$M_{\odot}$.  These clusters are not evenly distributed
but themselves clustered in super-clusters spanning regions of
several hundred pc in projected diameter and have concentrated
cores, i.e. the cluster-density in the centre of the
super-cluster is higher than in the outer parts.  The richness of
the super-clusters spans from groups of only a few to
super-clusters containing hundreds of new star clusters.  In
other systems like Arp~245, Duc et al.\ (2000) find a bound
stellar and gaseous object at the tip of the tidal tail.  (Their
numerical models show that the system is seen at approximately
100~Myr after the closest approach of the interacting pair.)
This so-called tidal dwarf galaxy (TDG) contains old and new
stellar material, a lot of gas and also undissolved massive
stellar clusters.  Systems like the merger remnant NGC~7252
(Miller et al.~1997) show a distribution of young massive star
clusters which has the same age as the interaction ($\approx
700$~Myr).  All these observations show that galaxy interactions
lead to the formation of new massive and compact star clusters.

Finally, galaxies like our Milky Way host several dwarf galaxies
(e.g. dSph or dE) (Mateo 1999) with high specific globular
cluster frequencies (Grebel 2000) and globular clusters in their
tidal streams. 

With this project we investigate the future fate of clusters of
young massive star clusters.  According to Kroupa (1998) it is
possible that such configurations merge and build a dwarf
galaxy.  Therefore we simulate compact super-clusters in the
tidal field of a host-galaxy and investigate the influence of
orbital and internal parameters on the rate and timescale of the
merging process (i.e. how fast the single clusters merge and how
many star clusters are able to survive this process).  In
addition the properties of the resulting merger object and its
dynamical evolution are studied.  A detailed description of the
properties of the merger objects will be given in a follow up
paper. 

\section{The Simulations}
\label{sec:setup}

The simulations are performed with the particle-mesh code {\sc
  Superbox} (Fellhauer et al. 2000).  In  {\sc Superbox}
densities are derived on Cartesian grids using the
nearest-grid-point scheme.  From these density arrays the
potential is calculated via a fast Fourier-transformation.  The
particles are then integrated forward in time using a fixed
time-step Leap-Frog algorithm.  {\sc Superbox} has a hierarchical 
grid architecture which includes for each object two levels of
high-resolution sub-grids. These sub-grids stay focused on the
objects and travel with them through the simulation space,
providing high resolution at the places of interest (in this case
the super-cluster and the single clusters within).   

The massive star clusters are simulated as Plummer-spheres
containing 100,000 particles each, having a Plummer-radius of
$R_{\rm pl} = 6$~pc and a cutoff radius $R_{\rm cut} = 30$~pc.
Each cluster has a total mass of $M_{\rm cl} = 10^{6}\ {\rm
  M}_{\odot}$ and a crossing time of $1.4$~Myr.  

The super-cluster is also modelled as a Plummer distribution made 
up of $N_{0}$ star clusters described above, has a Plummer-radius
$R_{\rm pl}^{\rm sc}$ and cutoff radius $R_{\rm cut}^{\rm sc} = 6
R_{\rm pl}^{\rm sc}$.  The Plummer-radius of the super-cluster
has values of 50, 75, 150 and 300~pc.  In this project the number
of clusters is kept constant at $N_{0} = 20$, which is a typical
number of star clusters found in these super-clusters.  $N_{0}$
was chosen small enough to get results with a considerable amount
of CPU-time.  Higher values of $N_{0}$ will be dealt with using
the newly available parallel version of {\sc Superbox} in the
near future.  In our calculations the super-clusters have an
initial velocity-distribution according to the
Plummer-distribution we gave them (i.e. they are initially in
virial equilibrium).  

The super-cluster orbits through the external potential of a
parent galaxy, which is given by 
\begin{eqnarray}
  \label{eq:galpot}
  \Phi(r) & = & \frac{1}{2} \ v_{\rm circ}^{2} \cdot \ln \left
  ( R_{\rm gal}^{2} + r^{2} \right) 
\end{eqnarray}
with $R_{\rm gal} = 4$~kpc and $v_{\rm circ} = 220$~km/s. We
refer the reader to Kroupa (1998), who describes the isolated
case.  The centre of the super-cluster moves on a circular orbit
at distance $D$ around the centre of the galaxy.  The distance
$D$ from the galactic centre is varied to be 5, 10, 20, 30, 50
and 100~kpc.  

The tidal radius $R_{t}$ of the super-cluster depends mainly on
$D$, but has also a low dependency on $R_{\rm pl}^{\rm sc}$.
$R_{t}$ lies at the local maxima of $\Phi_{\rm eff}$.  It can be
derived numerically by setting $\partial \Phi_{\rm eff} /
\partial r = 0$ where $\Phi_{\rm eff}$ is  
\begin{eqnarray}
  \label{eq:phieff}
  \Phi_{\rm eff}(r) & = & \frac{1}{2} \ v_{\rm circ}^{2} \cdot
      \ln \left( R_{\rm gal}^{2} + r^{2} \right) \ - \
      \frac{GM}{R_{\rm pl}^{\rm sc}} \cdot \left( 1 + \left
      ( \frac{r-D}{R_{\rm pl}^{\rm sc}}\right)^{2}
      \right)^{-1/2} \\  
      &   & - \ \frac{1}{2} \left( \frac{v_{\rm
      circ.orb.}}{D} \cdot r \right)^{2}. \nonumber 
\end{eqnarray}

The grids in this project are chosen to have $64^{3}$ mesh-points 
with the following sizes:  
\begin{itemize}
\item The innermost grids cover single star clusters and have
  sizes ($2 \cdot R_{\rm core}$) of 60~pc.  This gives a
  resolution of 1~pc per cell. 
\item The medium grids have sizes ($2 \cdot R_{\rm out}$)
  approximately equal to the cut-off radius of the super-cluster
  ($R_{\rm cut}^{\rm sc}$) to ensure that every star cluster is
  in the range of the medium grid of every other cluster.  This
  means the medium grids have the sizes shown in
  Table~\ref{tab:grid} 
  \begin{table}[h!]
      \caption{Grid sizes and resolutions per cell of the medium
        grids.} 
      \label{tab:grid}
    \begin{center}
      \hspace*{-7cm}
      \begin{tabular}[c!]{ccc}\hline
        $R_{\rm cut}^{\rm sc}$ & $2*R_{\rm out}$ & resolution \\
        \hline 
        300 pc & 600 pc & 10.0 pc\\
        450 pc & 1,000 pc & 16.7 pc\\
        900 pc & 2,000 pc & 33.3 pc \\
        1,800 pc & 3,000 pc & 50.0 pc \\ \hline
      \end{tabular}
    \end{center}
  \end{table}
\item The outermost grid (size: $2 \cdot R_{\rm system}$) covers
  the orbit of the super-cluster around the galactic centre.
  This means $2 \cdot R_{\rm system}$ is chosen to be 10~kpc
  larger than $2 \cdot D$, where $D$ is the distance of the
  super-cluster from the galactic centre. 
\end{itemize}

The two-dimensional parameter-space of our simulations ($R_{\rm
  pl}^{\rm sc}$ and $D$) can also be described with two
dimensionless variables, namely
\begin{eqnarray}
  \label{eq:dimless-parameter}
  \alpha & = & R_{\rm pl} / R_{\rm pl}^{\rm sc} \hspace*{2cm}
  \beta  \ = \ R_{\rm cut}^{\rm sc} / R_{t} 
\end{eqnarray}
$\alpha$ describes how densely the super-cluster is filled with
star clusters.  $\beta$ describes the strength of the tidal
forces acting on the super-cluster.

Table~\ref{tab:parameter} lists the dimensionless parameters
$\alpha$ and $\beta$ for the different choices of the
physical quantities $R_{\rm pl}^{\rm sc}$ and~$D$.
\begin{table}[h!]
  \begin{center}
    \caption{Dimensionless parameters as function of the physical
      quantities.} 
    \label{tab:parameter}
    \begin{tabular}[h!]{cccccccccc} \hline
      $\alpha$ & \multicolumn{4}{c}{$R_{\rm pl}^{\rm sc}$ [pc]} 
      & $\beta$ & \multicolumn{4}{c}{$R_{\rm pl}^{\rm sc}$ [pc]} 
      \\ 
      $D$ [kpc] & 50 & 75 & 150 & 300 & $D$ [kpc] & 50 & 75 & 150
      & 300 \\ \hline 
      5 & 0.12 & 0.08 & 0.04 &      & 5 & 0.775 & 1.178 & 2.500 &
      \\ 
      10 & 0.12 & 0.08 & 0.04 &      & 10 & 0.617 & 0.932 & 1.935
      & \\ 
      20 & 0.12 & 0.08 & 0.04 & 0.02 & 20 & 0.419 & 0.630 & 1.282
      & 2.761 \\ 
      30 & 0.12 & 0.08 & 0.04 & 0.02 & 30 & 0.323 & 0.486 & 0.981
      & 2.048 \\ 
      50 & 0.12 & 0.08 & 0.04 & 0.02 & 50 & 0.232 & 0.348 & 0.700
      & 1.431 \\ 
      100 &     &      &      & 0.02 &100 &        &      &
      & 0.89 \\ \hline
    \end{tabular}
  \end{center}
\end{table}

For each combination of ($\alpha$, $\beta$) several (3--6) random
realisations are performed.  Results discussed later for one pair
of the parameter set are computed mean values out of the
different simulations.  Our study covers more than the
observed ranges of parameters to show the influences of the
different parameters more clearly.

This is a first theoretical and numerical approach to investigate
the fate of the super-clusters.  Although, in nature, it is
unlikely that two star clusters form overlapping but with
uncorrelated velocities, we do not reject random number placements
which put two clusters in a distance where they
already overlap.  Our simulations do not include gas dynamics,
therefore they are valid after the expulsion of gas from
the star clusters.  By that time, clusters in the centre could
already overlap each other.   

\section{Results}
\label{sec:result}

The number of star clusters in the super-cluster decreases due to
two concurrent processes.  The first and also the most important
one is the merging process.  The second one is the escape of star
clusters from the super-cluster.  Escape plays an important role
only on long timescales or if the super-cluster is larger than
its tidal radius (i.e. $\beta>1$). 
\begin{eqnarray}
  \label{eq:nofc}
  N(t) & = & N_{0} - n_{\rm m}(t) - n_{\rm esc}(t)
\end{eqnarray}
where $n_{\rm m}$ is the number of merged clusters and $n_{\rm
  esc}$ the number of escaped clusters.

\subsection{Merging Timescales}
\label{sec:timescale}

In our simulations the timescale of the merger process is very
short.  Most cluster merge within the first few crossing times of
the super-cluster, forming a dense and spherical merger object in
the centre of the super-cluster.  

To determine the timescale of this merger process we first have
a look at simulations with $\beta<1.0$.  With this restriction we
can neglect the number of escaping clusters (i.e. $n_{\rm esc}(t)
\equiv 0$).

Then we take the following ansatz for the decrease of the number
of clusters.  If a star cluster makes one crossing through the
super-cluster the chance to meet and merge with another cluster
is the area covered by the cross-sections of all other clusters
divided by the area of the super-cluster.  We therefore define
the merger-rate $R$ (number of merger events per dimensionless
time unit) as the ratio between the cross-sections of all
($N$) star clusters ($A_{\rm hit}$) travelling through the
super-cluster and the area of the super-cluster $A_{\rm sc}$.  
The merger rate $R$ per $T_{\rm cr}^{\rm sc}$ (crossing time of 
the super-cluster) should therefore be given by 
\begin{eqnarray}
  \label{eq:ansatz1}
   R \ = \ - \frac{{\rm d}N}{{\rm d}\tau} & = & N \cdot
       \frac{A_{\rm hit}}{A_{\rm sc}} 
\end{eqnarray}
with 
\begin{eqnarray}
  \label{eq:tau}
  \tau & = & \frac{t}{T_{\rm cr}^{\rm sc}}
\end{eqnarray}
being a dimensionless time and 
\begin{eqnarray}
  \label{eq:scarea}
  A_{\rm sc} & = & \pi \cdot (R_{\rm cut, eff}^{\rm sc})^{2} \\
  & = & \pi \cdot (\gamma \cdot R_{\rm pl}^{\rm sc})^{2}
  \nonumber 
\end{eqnarray}
is the projected area of the super-cluster with a mean radius
that includes all clusters.  This radius is smaller than $R_{\rm
  cut}^{\rm sc}$ due to the fact that the super-clusters contain
only a limited number of clusters. In our simulations we find
that $\gamma$ is given by $\gamma = 3.8 \pm 0.1$.

\begin{figure}[h!]
  \begin{center}
    \epsfxsize=5cm
    \epsfysize=5cm
    \epsffile{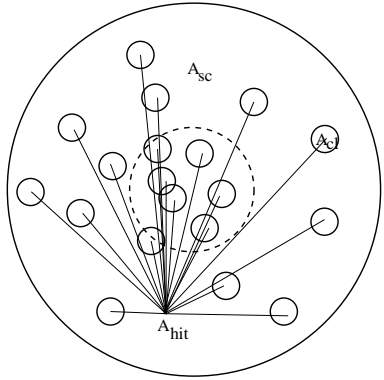}
    \caption{Schematic view of the first ansatz to determine the
      merging timescales.} 
    \label{fig:an1}
  \end{center}
\end{figure}

For the cross-sections of all clusters we take the ansatz (see
also Fig.~\ref{fig:an1})
\begin{eqnarray}
  \label{eq:mergrateareas}
  A_{\rm hit} & = & (N-1) \cdot A_{\rm cl} \hspace*{1cm} {\rm
  with} \\
  A_{\rm cl} & = & \pi \cdot r_{\rm m}^{2} \nonumber
\end{eqnarray}

Every star cluster sees the cross-section $A_{\rm cl}$ of $N-1$ 
other clusters and $r_{\rm m}$ being the maximum distance at
closest approach which leads to a merger afterwards.  $R$ is then
proportional to $N^{2}$ and the number of clusters $N(\tau)$
should decrease with time proportional to $1/(1+k\tau)$. 

\begin{figure}[h!]
  \begin{center}
    \epsfxsize=12cm
    \epsfysize=08cm
    \epsffile{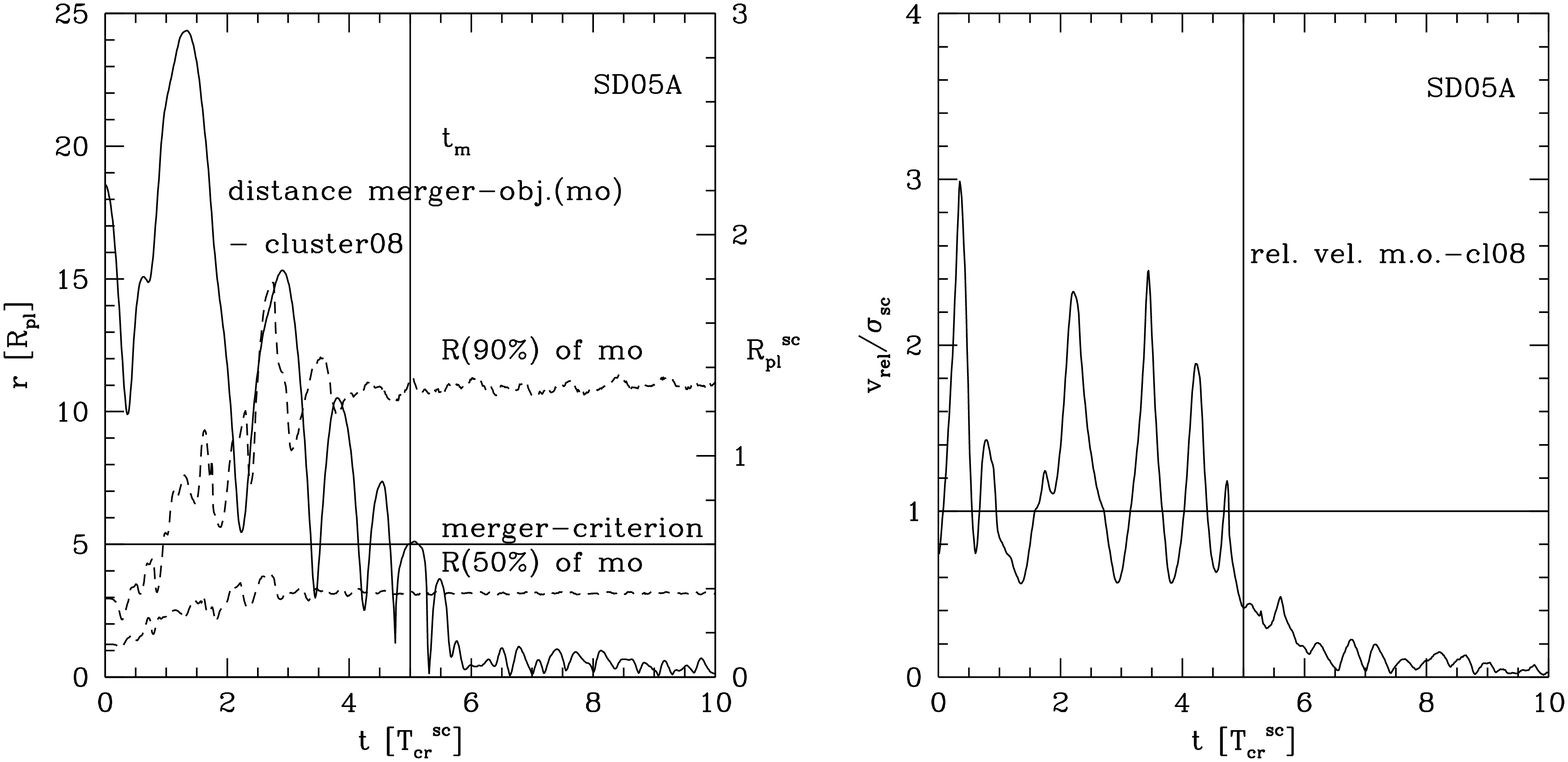}
    \caption{A typical example for the merging of a star cluster
    with the merger object.  Left panel: Solid line shows the
    distance between the merger object and the star cluster.
    Dashed lines show the half-mass- and the 90\%-radii of the
    merger object. Horizontal line marks the maximum distance for
    the merger-criterion -- if the distance of two objects stays
    less than this for the rest of the simulation the two objects
    are assumed to be merged; vertical line shows the adopted
    merger-time $t_{\rm m}$.  Right panel: Ratio between the
    relative velocity $v_{\rm rel}$ of the cluster and the
    merger object and the velocity dispersion $\sigma_{\rm sc}$
    of the super-cluster.}  
    \label{fig:enc05}
  \end{center}
\end{figure}

To determine $A_{\rm cl}$ we first have to check how the average
encounter between two star clusters (or a star cluster and the
merger object) looks like.  Therefore we have to check the
relative velocities and separations at the point of closest
approach, which afterwards leads to the merging of the two star
clusters.  Fig.~\ref{fig:enc05} shows the relative separation
between a particular star cluster and the merger object.  One can
clearly see that the two objects are separated at their first
encounter, implying the distance at closest approach is larger
than the mean square radius of a single cluster.  Their relative
velocity is larger than the average velocity inside the
super-cluster but less than a factor of three.  Because even
head-on encounters do not lead to merging if the encounter
velocity lies above a certain treshhold, we follow a simple
criterion extracted from Gerhard \& Fall (1983 their Fig.~1)
\begin{eqnarray}
  \label{eq:velcheck}
  2 \cdot \sigma_{\rm sc}^{2} & \leq & \frac{GM_{\rm cl}}
  {R_{\rm pl}}          
\end{eqnarray}
where $\sqrt{2}\sigma_{\rm sc}$ is the typical relative velocity
of a pair of star clusters.  This equation holds for $\alpha \leq
0.085$.  For $\alpha > 0.085$ we argue below that in our parameter 
range we already start with a merged object in the centre of the
super-cluster, and therefore $M_{\rm cl}$ has to be replaced by
$\nu M_{\rm cl}$ where $\nu$ is the number of merged central
clusters.  With $\nu \geq 2$ and increasing with increasing
$\alpha$, Eq.~\ref{eq:velcheck} holds for all values of $\alpha$
in our parameter space. 

In this paper, we define two clusters as merged if their mutual
distance stays smaller then five Plummer-radii $R_{\rm pl}$ for
the rest of the simulation.  We found that $5\,R_{\rm pl}$ is a
good compromise between declaring no cluster as merged (the
merger criterion being chosen too small) and merging all clusters
right form the beginning (their mean distance being smaller than
the merger criterion).  Smaller merger radii have problems in the
late stages of the simulation, where the centres of density of
dissolved clusters are very hard to determine and could be found
off-centre of the extended merger object.  If the merger radius
is chosen too small, these clusters would not be counted as
merged.  An energy criterion has to be handled with care,
because clusters may be bound to the merger object without
merging but staying on a circular orbit and decaying on the
dynamical friction timescale. 

The next step is to approximate the energy gain of the clusters
due to the passage.  Aguilar \& White (1985) showed that the
total energy exchange in the tidal approximation given by Spitzer
(1958),
\begin{eqnarray}
  \label{eq:spitzer}
  \Delta E & = & \frac{1}{2}\ M_{\rm cl} \left( \frac{2GM_{\rm
  cl}} {r_{\rm p}^{2}v_{\rm p}}\right)^{2}\ \frac{2}{3} r_{\rm
  c}^{2}, 
\end{eqnarray}
gives reasonable results if $r_{\rm p} \geq 5 r_{\rm c}$.
In this formula $r_{\rm c}$ denotes the mean square radius of a
single cluster.  With a cut-off of $6R_{\rm pl}$ we calculate the
mean square radius of a Plummer-sphere as approximately $1.5
R_{\rm pl}$ which gives $r_{\rm c}^{2} \approx 2.3 R_{\rm pl}^{2}$.
Fig.~\ref{fig:enc05} shows that the two objects pass each other
for the first time at a distance which holds for the criterion of
Aguilar \& White.  $r_{\rm p}$ and $v_{\rm p}$ are the distance
and the velocity at the point of closest approach.  If we assume
that $v_{\rm p}$ is equal to the mean relative velocity of two
clusters in the super-cluster, $\sqrt{2} \sigma_{\rm sc}$, i.e.\
we neglect gravitational focusing and the acceleration of the
star clusters firstly and taken into account that both clusters
are able to gain energy we get 
\begin{eqnarray}
  \label{eq:spitzer2}
  \Delta E & = & 2 \frac{4 \cdot 2.3}{3 \cdot 2} \frac{G^{2} M_{\rm
  cl}^{3} R_{\rm pl}^{2}}{r_{\rm p}^{4} \sigma_{\rm sc}^{2}}.
\end{eqnarray}
To obtain the critical impact parameter which leads to a merger,
we set the energy exchange equal to the orbital energy of the two
clusters in the super-cluster: $\Delta E = \frac{1}{4}\ M_{\rm
  cl}\ (\sqrt{2}\sigma_{\rm sc})^{2}$.  Inserting for
$\sigma_{\rm sc}$ the mean velocity of ``particles'' in a
Plummer-sphere 
\begin{eqnarray}
  \label{eq:plvel}
  \sigma_{\rm sc}^{2} = \frac{3\pi}{32} \frac{GM_{\rm sc}}{R_{\rm
  pl}^{\rm sc}}
\end{eqnarray}
with $M_{\rm sc} = N_{0} M_{\rm cl}$ and using the definition of
$\alpha$ from Eq.~\ref{eq:dimless-parameter}, one obtains for
$r_{\rm p}$ 
\begin{eqnarray}
  \label{eq:infrad}
  r_{\rm p} & = &
  \left( \frac{8\cdot32^{2}\cdot2.3} {3(3\pi)^{2}N_{0}^{2}}
  \right)^{1/4} \cdot \sqrt{\alpha} \cdot R_{\rm pl}^{\rm sc} \
  \approx \ 0.65 \sqrt{\alpha} \cdot R_{\rm pl}^{\rm sc}. 
\end{eqnarray}

\begin{figure}[h!]
  \begin{center}
    \epsfxsize=12cm
    \epsfysize=12cm
    \epsffile{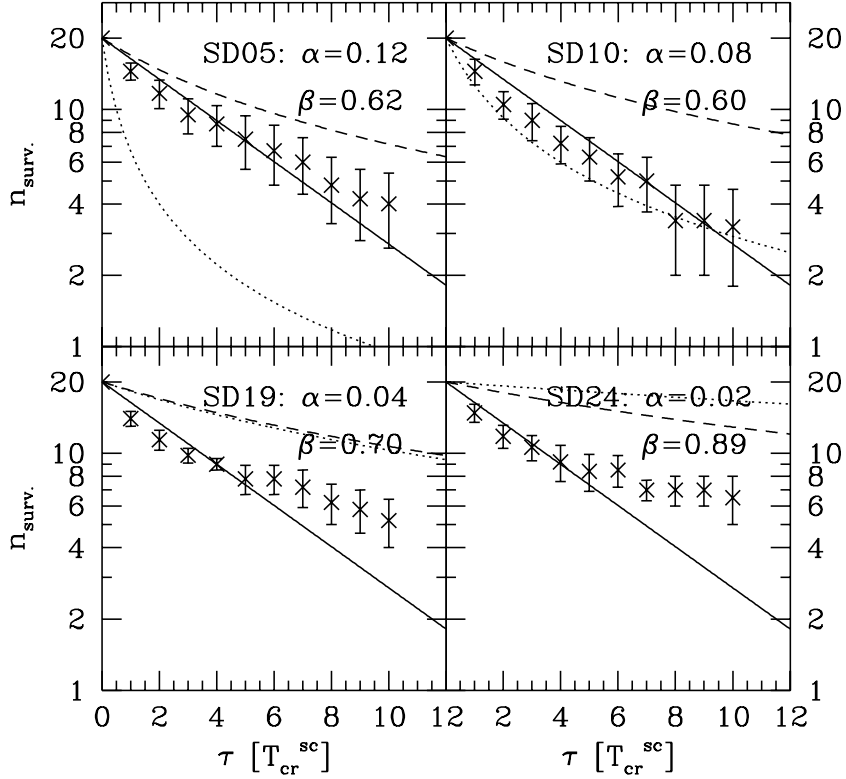}
    \caption{Number of remaining objects vs.\ time measured in 
      crossing times of the super-cluster for different values of 
      the parameter $\alpha$.  $\beta$ is kept smaller than
      $1.0$.  Lines show theoretical curves.  Solid line
      corresponds to an exponential decrease, dashed line shows
      the $1/(1+k\tau)$-decrease which depends on $\alpha$.
      Dotted lines  rely on the Aarseth \& Fall criterion
      (Eq.~\ref{eq:af3}).}  
    \label{fig:mtime}
  \end{center}
\end{figure}

In this formula we have not taken into account that the clusters
are gravitationally focused.  Inserting Spitzers (1987; eq.~6-15)
formula for gravitational focusing, 
\begin{eqnarray}
  \label{eq:gravfocus}
  r_{\rm m} & = & r_{\rm p} \sqrt{ 1.0 + \frac{4GM_{\rm cl}}
    {r_{\rm p} (\sqrt{2}\sigma_{\rm sc})^{2}}},
\end{eqnarray}
we obtain
\begin{eqnarray}
  \label{eq:infrad2}
  r_{\rm m} & = & 0.65 \cdot \sqrt{\alpha} \cdot \sqrt{ 1.0 +
  \frac{0.52}{\sqrt{\alpha}}} \cdot R_{\rm pl}^{\rm sc}.
\end{eqnarray}
Given the values of $\alpha$ of our simulations we derive the
values for $r_{\rm m}$ given in Table~\ref{tab:infrad}.
\begin{table}[h!]
  \begin{center}
    \hspace*{-4.cm}
    \begin{tabular}[h!]{c|cccccc}
      $\alpha$ & 0.24 & 0.12 & 0.08 & 0.04 & 0.02 & 0.006 \\ \hline
      $r_{\rm m}$ [$R_{\rm pl}^{\rm sc}$] & 0.46 & 0.36 & 0.31 &
      0.25 & 0.20 & 0.14 \\
      $r_{\rm m}$ [$R_{\rm pl}$] & 1.91 & 2.97 & 3.88 & 6.18 &
      9.96 & 23.4 \\
      $\delta(\alpha) \cdot N_{0}$ & 0.29 & 0.18 & 0.13 & 0.087 &
      0.055 & 0.027  
    \end{tabular}
    \caption{Merger radius $r_{\rm m}$ in units of $R_{\rm
      pl}^{\rm sc}$ and $R_{\rm pl}$ and proportionality factor
      $\delta N_{0}$ for the different $\alpha$-values.}
    \label{tab:infrad}
  \end{center}
\end{table}
The merger rate $R$ is then
\begin{eqnarray}
  \label{eq:mr}
  R & = & N \cdot (N-1) \cdot \frac{(0.65)^{2} \alpha (1 +
  \frac{0.52} {\sqrt{\alpha}})} {\gamma^{2}} \ \approx \
  \delta(\alpha) \cdot N^{2}. 
\end{eqnarray}
Solving for $R=-{\rm d}N/{\rm d}\tau$ gives 
\begin{eqnarray}
  \label{eq:ndecr1}
  N(\tau) & = & N_{0} \cdot \frac{1}{1+\delta(\alpha) N_{0}
  \tau}, 
\end{eqnarray}
with the values for $\delta(\alpha)\cdot N_{0}$ shown in
Table~\ref{tab:infrad}.  This dependency is plotted as dashed
lines in Figs.~\ref{fig:mtime} \&~\ref{fig:ctime}. 

On the other hand, using instead the merger criterion by Aarseth
\&  Fall (1980) as described in Gerhard \& Fall (1983),
\begin{eqnarray}
   \label{eq:af1}
   \frac{r_{\rm p}^{2}} {[4R_{\rm h}]^{2}} + \frac{v_{\rm p}^{2}}
   {[1.16 v_{\rm esc}(p)]^{2}} & \leq & 1,
\end{eqnarray}
(their Eq.~4) where
\begin{eqnarray}
   \label{eq:af2}
   v_{\rm esc}(p)^{2} & = & \frac{4GM_{\rm cl}}{(r_{\rm p}^2 + 
   2R_{\rm pl}^{2})^{1/2}},
\end{eqnarray}
(their Eq.~5) with $R_{\rm h}$ being the half-mass radius of a
Plummer-sphere ($\approx 1.3 R_{\rm pl}$) and $v_{\rm p} =
\sqrt{2}\sigma_{\rm sc}$ we obtain
\begin{eqnarray}
   \label{eq:af3}
   r_{\rm p} & \leq & \left( 27.04 + 67.6 \sqrt{671.75 \alpha^{2}
   + 22.31} \cdot \alpha + 1754.79 \cdot \alpha^{2} \right)^{1/2} \\
   & & \hspace{8cm} \cdot \ \alpha \ R_{\rm pl}^{\rm sc} \nonumber
\end{eqnarray}
Using $r_{\rm p}$ from this expression instead of $r_{\rm m}$ in our 
Eq.~\ref{eq:mergrateareas} we obtain the dotted curves in
Figs.~\ref{fig:mtime} \&~\ref{fig:ctime}.  We note that this
ansatz does not lead to a good description of our results.  It
leads to an $\alpha$ dependency of $r_{\rm p} \propto
\alpha^{2}$, which is steeper than we find in our simulations.  Also
the merger theory of Makino \& Hut (1997) does not agree with our
results.  They find a dependency on $\alpha$ with a power of 1.5,
which does not fit our data.  In our simulations we find a very
weak dependency of the merging timescales on the parameter
$\alpha$.  For our main parameter range ($\alpha = 0.02 - 0.12$)
there is even no visible dependency.  The simulations show a
clear exponential decrease of the number of clusters with
dimensionless time $\tau$ independent of the choice of $\alpha$
(see Fig.~\ref{fig:mtime}).  Therefore, a merger theory depending
on the merging of single clusters must be wrong.

The mean projected distance of clusters inside the innermost
Plummer radius of the super-cluster depends only on the choice of
$N_{0}$ and is given by 
\begin{eqnarray}
  \label{eq:meandist}
  d_{\rm mean}(1R_{\rm pl}^{\rm sc}) & = & \kappa \cdot
  \sqrt{\frac{\pi (R_{\rm pl}^{\rm sc})^{2}} {N_{\rm proj}}}.
\end{eqnarray}
Numerical simulation shows that $\kappa$ is given by 0.53.  Since
half the star clusters lie within one projected Plummer radius,
we get for our choice of $N_{0}=20 \rightarrow N_{\rm proj}=10$  
\begin{eqnarray}
  \label{eq:meandist2}
  d_{\rm mean}(1R_{\rm pl}^{\rm sc}) & \approx & 0.30 R_{\rm
  pl}^{\rm sc}. 
\end{eqnarray}
Comparing this value with the merger radii $r_{\rm m}$ for our
choices of $\alpha$ (see Table~\ref{tab:infrad}) one can see that
in the centre of the super-cluster the mean distance between two
clusters (within the innermost Plummer-radius $R_{\rm pl}^{\rm
  sc}$) is comparable the merger radius for most $\alpha$.
In the centre the clusters are not able to separate from each
other right from the beginning and the merging of them should
therefore happen very quickly within one or two crossing times of
the super-cluster, especially for high $\alpha$.  

\begin{figure}[h!]
  \begin{center}
    \epsfxsize=5cm
    \epsfysize=5cm
    \epsffile{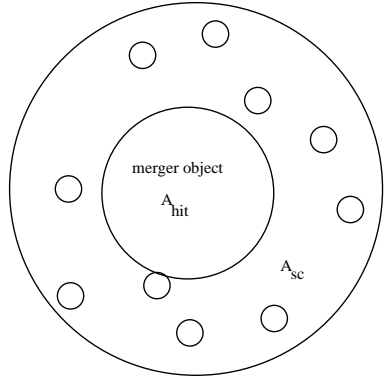}
    \caption{Schematic view of the new ansatz to determine the
    merging timescales.} 
    \label{fig:an2}
  \end{center}
\end{figure}

In this case we have to deal with a big merger object covering
the central area of the super-cluster.  Merging happens
preferably with this central object.  Therefore, our ansatz for
$A_{\rm hit}$ in Eq.~\ref{eq:mergrateareas} fails.  Instead we use
the following ansatz shown in Fig.~\ref{fig:an2}.
The merging cross-section $A_{\rm hit}$ is now a fraction
$\epsilon$ of the area of the whole super-cluster $A_{\rm sc}$
\begin{eqnarray}
  \label{eq:ansatz2}
  A_{\rm hit} & = & \epsilon A_{\rm sc}.
\end{eqnarray}
which has to be determined.  Inserting in Eq.~\ref{eq:ansatz1}
($R=-\rm{d}N/\rm{d}\tau = \epsilon N$) and integrating leads to
\begin{eqnarray}
  \label{eq:ndecr2}
  N(\tau) & = & N_{0} \cdot \exp(-\epsilon \tau).
\end{eqnarray}
The exponential decrease, according to Eq.~\ref{eq:ndecr2}, is
plotted in Fig.~\ref{fig:mtime} as the solid lines.  There one
sees clearly that the number of clusters first decreases
exponentially.  One should also notice that the merging timescale
is now independent of the choice of $\alpha$. 

After all clusters travelling through the centre have
merged with the central merger object, the further decrease of
$N$ with time levels off.  This is due to the fact that even if
$\beta < 1.0$ there is the chance that some clusters escape and
as a second effect there are clusters on rather circular orbits
which do not travel through the central area.  In low $\alpha$
cases ($\alpha = 0.04, 0.02$) these clusters do not merge but
their orbits will decay on timescales of the dynamical friction.

\begin{figure}[h!]
  \begin{center}
    \epsfxsize=12cm
    \epsfysize=08cm
    \epsffile{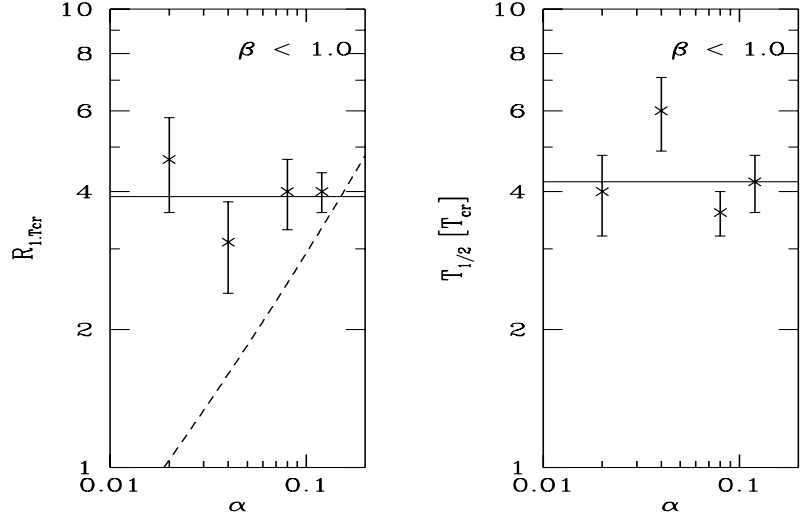}
    \caption{$R_{1.T_{\rm cr}^{\rm sc}}$ and $T_{1/2}$ as
    function of $\alpha$ for all simulations with $\beta<1$.
    Solid lines show mean values from Eq.~\ref{eq:simulval} while 
    dotted line corresponds to the $\alpha$-dependent theory from 
    Eq.~\ref{eq:mr}.}  
    \label{fig:alpha}
  \end{center}
\end{figure}

To prove the independency on $\alpha$ we consider two
characteristic quantities, namely the mean merger rate within the
first crossing time ($R_{1.T_{\rm cr}^{\rm sc}}$) and the
half-life merging time ($T_{1/2}$), i.e. the dimensionless time
until $N_{0}/2$ of the clusters have merged.
Fig.~\ref{fig:alpha} shows the number of clusters merging within
the first crossing time, and the time it takes to merge half the
clusters.  All simulations with the same $\alpha$, and which have
$\beta<1$, are binned together.  As can be seen, there is no
dependency on $\alpha$. 

\begin{figure}[h!]
  \begin{center}
    \epsfxsize=12cm
    \epsfysize=12cm
    \epsffile{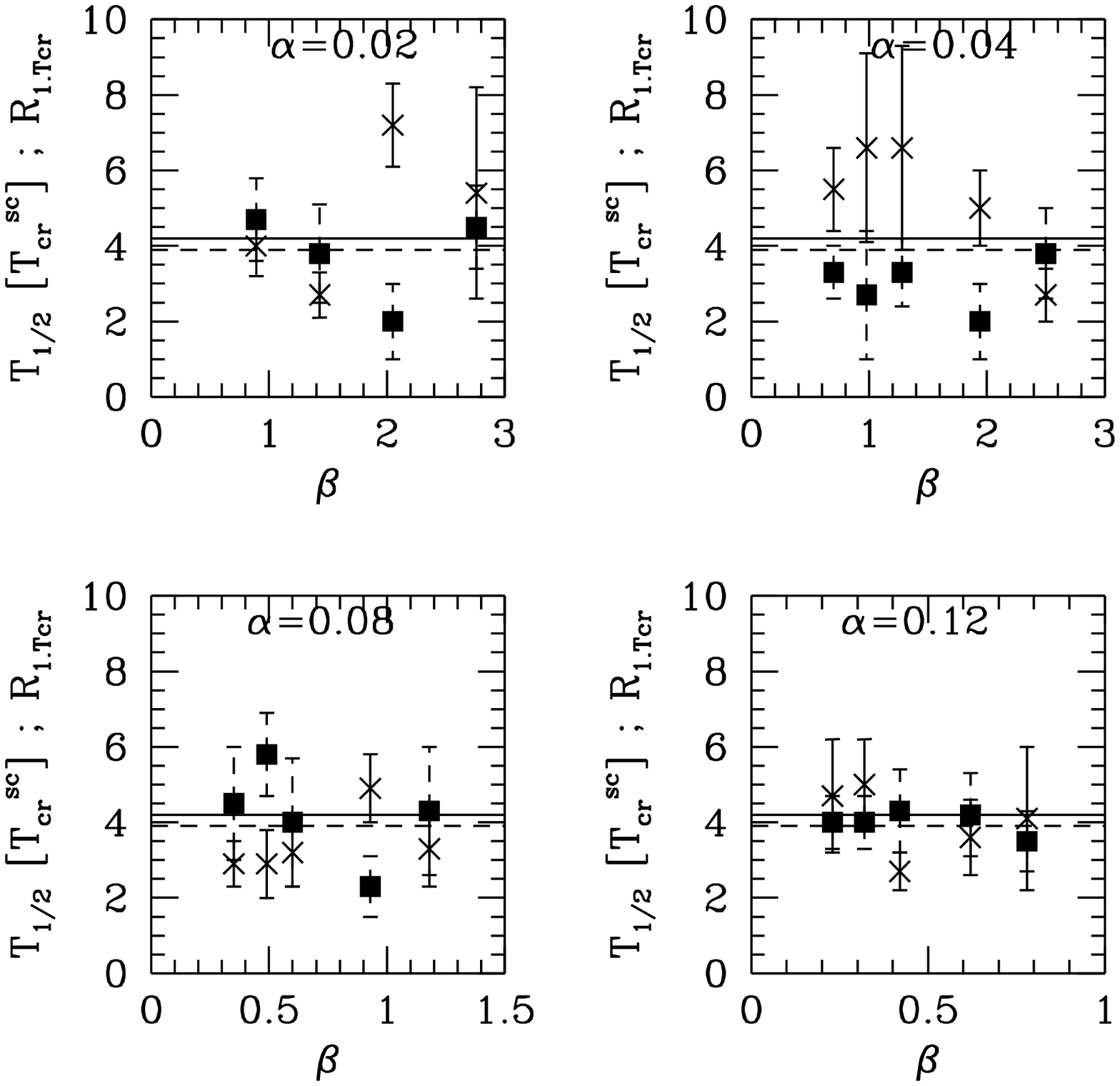}
    \caption{$R_{1.T_{\rm cr}^{\rm sc}}$ (boxes) and $T_{1/2}$
    (crosses) as a function of $\beta$ for the different choices 
    of $\alpha$.  Solid line is the mean value for $T_{1/2}$,
    dashed line for $R_{1.T_{\rm cr}^{\rm sc}}$ from
    Eq.~\ref{eq:simulval}.}  
    \label{fig:beta}
  \end{center}
\end{figure}

The mean merger rate within the first crossing time and the mean
half-life merging time derived from our simulations are given by
\begin{eqnarray}
  \label{eq:simulval}
  R_{1.T_{\rm cr}^{\rm sc}} & = & 3.90 \pm 0.32 \ [{\rm merger \
  events}/T_{\rm cr}^{\rm sc}]\\ 
  T_{1/2} & = & 4.18 \pm 0.35 \ [T_{\rm cr}^{\rm sc}]
  \nonumber 
\end{eqnarray}
are plotted in Fig.~\ref{fig:alpha} and are also displayed in
Fig.~\ref{fig:beta}.  Inserting this values in our ansatz
(Eq.~\ref{eq:ansatz2}) for the merging times gives  
\begin{eqnarray}
  \label{eq:epsilon}
  \epsilon & \approx & 0.2.   
\end{eqnarray}

Fig.~\ref{fig:beta} shows the mean values of $R_{1.T_{\rm
    cr}^{\rm sc}}$ and $T_{1/2}$ for all combinations of
($\alpha$, $\beta$).  It is clearly visible that there is no
dependency of the results on $\beta$ at all.  It therefore seems
that $\beta$ influences the number of clusters which are able to
merge rather than the timescale of the merger process. 

We expect to obtain a change in the initial behaviour of the
system if $r_{\rm m}$ falls well below the mean projected
distance of the clusters.  For our choice of $N_{0}$, this should
happen approximately for $\alpha \leq 0.02$.  Then our first
ansatz (Eq.~\ref{eq:ansatz1}) should represent the system.  On
the other hand if $\alpha$ becomes higher than $\approx$~0.2, the
mean distance of all clusters is smaller than the merger radius
-- all $N_{0}$ star clusters should then merge within one or two
crossing times.  To prove this behaviour we performed test
calculations without tidal field.  The results
(Fig.~\ref{fig:ctime}) show nicely the transition between these
three regimes.  The upper left panel is a simulation with $\alpha
= 0.24$ and one sees clearly that almost all clusters merge
within the first two crossing times.  In the lower right panel a
simulation with $\alpha = 0.006$ is displayed.  The number of
clusters decreases according to our $\alpha$-dependent theory.

\begin{figure}[h!]
  \begin{center}
    \epsfxsize=12cm
    \epsfysize=15cm
    \epsffile{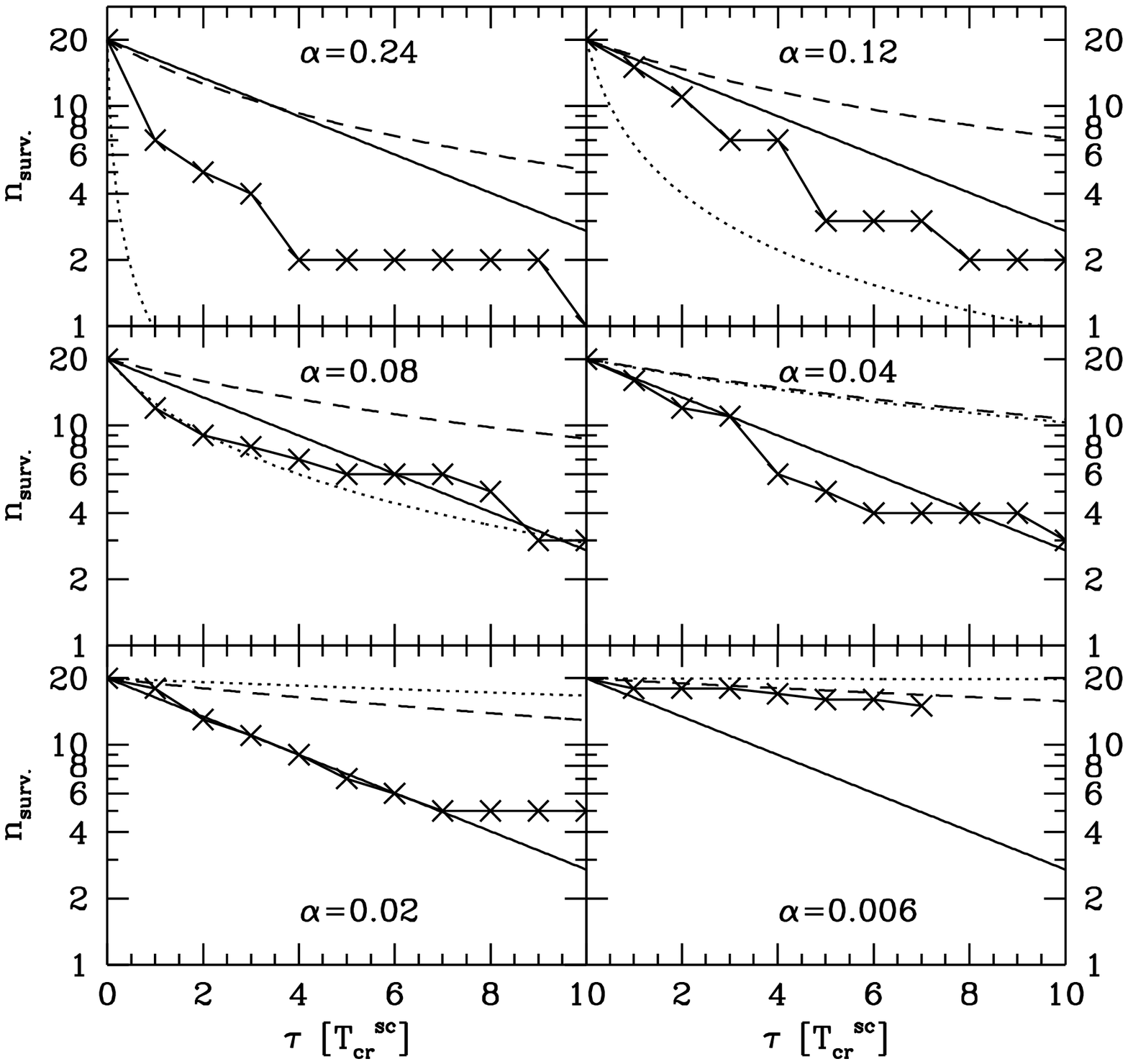}
    \caption{Same as Fig~\ref{fig:mtime} but without tidal field
      and more extended $\alpha$-range.  Errorbars are omitted
      because only one simulation for each $\alpha$-value was
      performed.}  
    \label{fig:ctime}
  \end{center}
\end{figure}

\subsection{Merger-Rates}
\label{sec:rate}

\begin{figure}[h!]
  \begin{center}
    \epsfxsize=12cm
    \epsfysize=10cm
    \epsffile{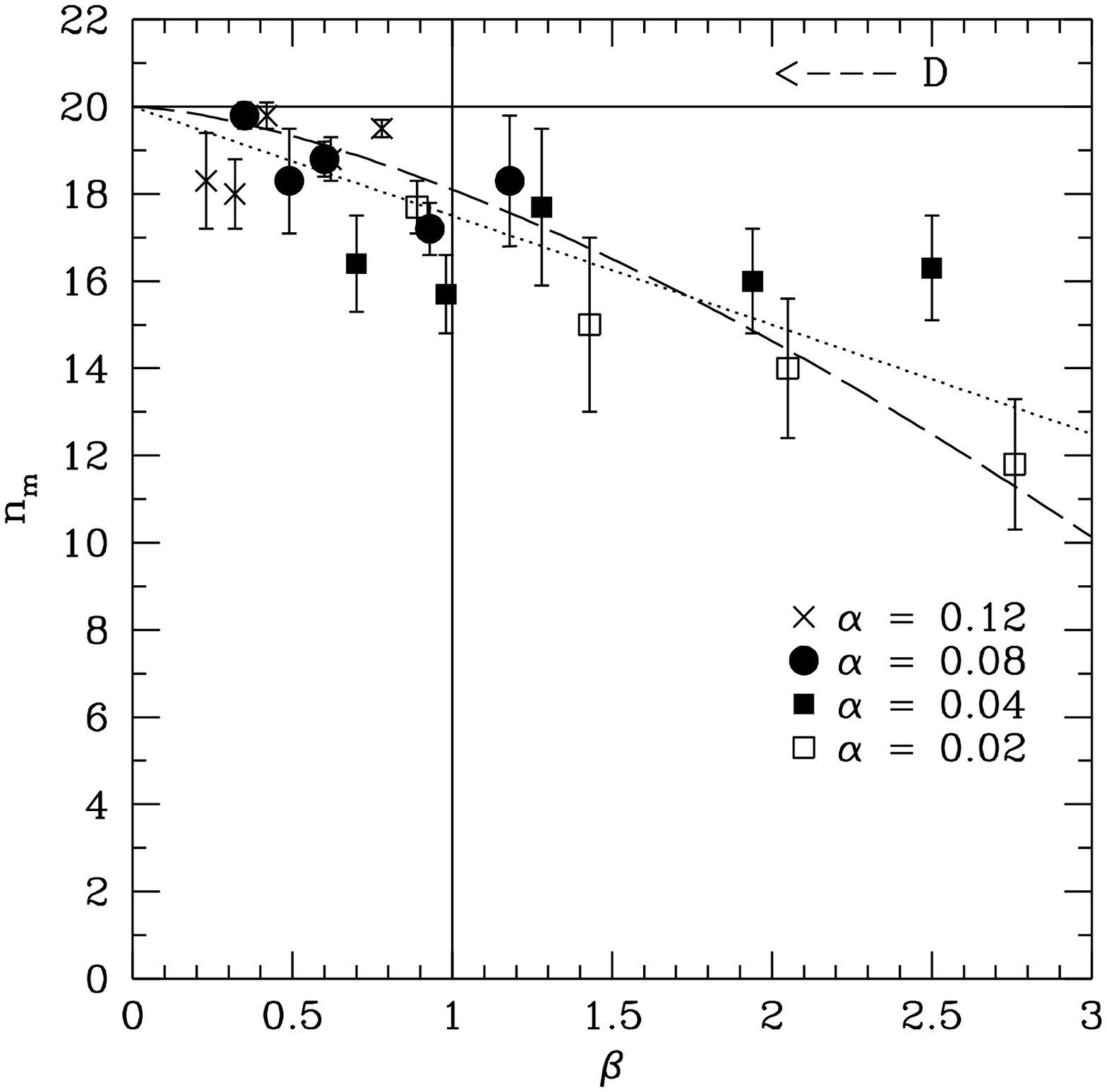}
    \caption{Number of merged clusters $n_{\rm m}$ vs.\ parameter
    $\beta$ for different values of $\alpha$.} 
    \label{fig:mrate}
  \end{center}
\end{figure}

The number of clusters, $n_{\rm m}$, which end up in the merger
object shows no significant dependency on $\alpha$ and $\beta$,
as long as the super-cluster configuration is well inside it's
tidal radius (i.e. $\beta<1.0$).  In this case almost all
clusters merge and only one or two clusters sometimes survive by
chance.  As one can see in Fig.~\ref{fig:mrate}, $n_{\rm m}$ is
close to $N_{0}=20$.  

This changes if $\beta$ becomes larger than 1.  There is a
significant drop in $n_{\rm m}$ which also shows a weak
$\alpha$-dependency.  

We interpret this result as follows:  We start with
Eq.~\ref{eq:nofc} and neglect the rare case of an escaping
merged cluster here.  If $\beta$ is small ($\beta < 1$) tidal
effects are not dominant, and the evolution of $N(t)$ is
determined by the merger processes alone as discussed before.  If 
$\beta > 1$ there is a trend that clusters can leave the
super-cluster before participating in the merger events.  The
number of escaping clusters depends on how many clusters
initially are outside the tidal radius and on the individual
velocities of such clusters.  With only 20 clusters initially
for the entire super-cluster any statistics of the subset of
escaping clusters is extremely poor.  Their number strongly
depends on the random numbers used for the initialisation of the
system.  Keeping in mind this poor statistical weight of our
data, we nevertheless identify two physically reasonable trends
in Fig.~\ref{fig:mrate}: first, the larger $\beta$, the smaller
the number of merging clusters, and second, this trend is more
pronounced for small $\alpha$.  This result is consistent with
the picture that strong tidal fields lead to more escapers
(i.e.\ less clusters available for merging); if, however, the
individual clusters are relatively extended (larger $\alpha$),
the merging competes with the escape; clusters on orbits of
potential escapers could be captured in the central regions by
merging with a higher probability. 

\begin{figure}[h!]
  \begin{center}
    \epsfxsize=12cm
    \epsfysize=10cm
    \epsffile{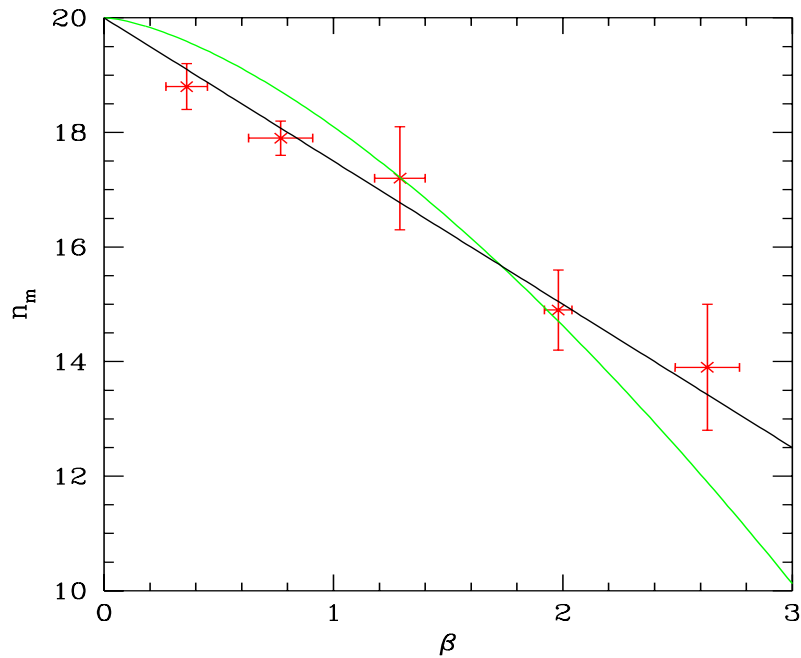}
    \caption{Number of merged cluster as function of $\beta$.
    Simulations are put together in 5 $\beta$-bins.  Straight
    line is the fitted linear decrease; curved line would be the 
    theoretical $\beta^{3/2}$-dependency as stated in Baumgardt
    (1998).}  
    \label{fig:betafit}
  \end{center}
\end{figure}
Binning the simulations in 5 $\beta$-bins, as shown in
Fig.~\ref{fig:betafit}, shows that the dependency on $\beta$ is
linear.  As best-fit for the data of our simulations we calculate
\begin{eqnarray}
  \label{eq:betafit}
  n_{\rm m} & = & 20 - (2.5 \pm 0.1) \cdot \beta .
\end{eqnarray}

To advance that kind of reasoning to an extreme, we could compare
our results with the findings of Baumgardt (1998), who models the
escaping stars from star clusters.  He proposes a
$\beta^{3/2}$-dependency.  Our results are consistent with
Baumgardt's results, in particular for $\alpha \rightarrow 0$ in
which case $N(\tau) \rightarrow N_{0}$ in Eq.~\ref{eq:ndecr1}
(where $n_{\rm esc}$ was neglected), so that $ N(t) \rightarrow
N_{0} - n_{\rm esc}(t)$ (Eq.~\ref{eq:nofc}). Note, however, that
our particle number  is very small compared to that work. 

\subsection{Building Up the Merger Objects}
\label{sec:mo}

The formation scenario of the merger object depends on the chosen
$\alpha$-value.  One finds that with high values of $\alpha$, one
has overlapping star clusters at the centre right from the
beginning, and the simulation already starts with a merger object
in the centre of the super-cluster.  With decreasing values of
$\alpha$, the merger-tree starts with the merging of clusters at
different positions in the super-cluster.  Afterwards, these
merger objects sink to the centre and merge together.  The exact
details of the merging depend however on the starting conditions,
and there can always be cases with high $\alpha$ behaving like
low $\alpha$ and vice versa.

\begin{figure}[h!]
  \begin{center}
    \epsfxsize=12cm
    \epsfysize=06cm
    \epsffile{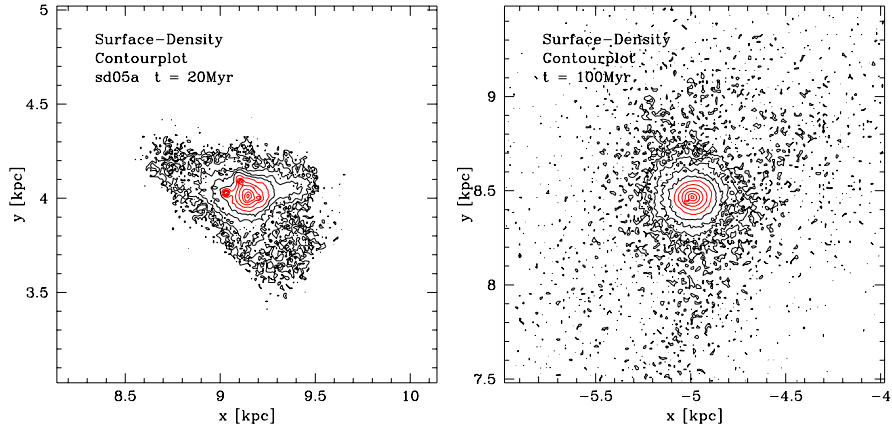}
    \caption{Surface-density contour plots (resolution: 
      $(10\ {\rm pc})^{2} $) of a simulation with
    $\alpha=0.12$ and $D=10$~kpc.  The crossing time of the
    super-cluster is 7.4~Myr.  High values of $\alpha$ correspond
    to compact super-clusters with short crossing times.}
    \label{fig:sd05a}
  \end{center}
\end{figure}

\begin{figure}[h!]
  \begin{center}
    \epsfxsize=12cm
    \epsfysize=12cm
    \epsffile{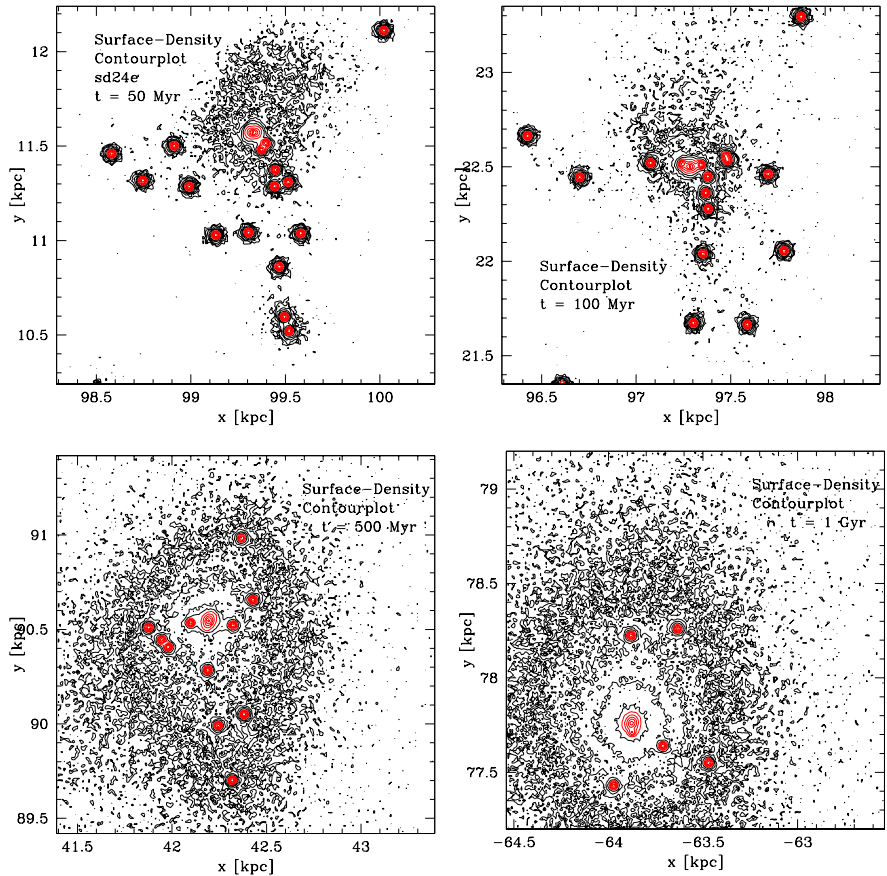}
    \caption{Surface-density contour plots (resolution 
      $(10\ {\rm pc})^{2} $) of a simulation with
    $\alpha=0.02$ and $D=100$~kpc.  The crossing time of the
    super-cluster is 108.4~Myr.} 
    \label{fig:sd24e}
  \end{center}
\end{figure}

Fig.~\ref{fig:sd05a} and~\ref{fig:sd24e} show some snapshots of
the evolution of clusters with high and low $\alpha$-values.   


Since the crossing-time of the super-cluster expressed in physical
units depends on its size and therefore on $\alpha$, the merger
rates, if expressed in Myrs, also depend on $\alpha$: Large
$\alpha$-values correspond to compact clusters which merge within
a few tens of a Myr, while small $\alpha$-values correspond to
extended clusters in which it can take up to 1 Gyr until all
clusters are finally merged. 

After 20~Myr almost every cluster has already fallen into the
merger object at the high $\alpha$ simulation
(Fig.~\ref{fig:sd05a}).  At $t=100$~Myr only the last single
cluster can be seen (as a disturbance in the contours) merging
with the main object.   

In the low $\alpha$ simulation (Fig.~\ref{fig:sd24e}) almost
every cluster is seen as an individual object, even after
100~Myr.  After 1~Gyr the merger object is surrounded by clusters
which are still in the process of merging.  These clusters may
account for a high specific cluster frequency of the merger
object.  Not shown in the last snapshot are the 2 escaping
clusters travelling on the same orbit around the galaxy as the
merger object.  

The merger objects show an exponential density distribution with
exponential scale-length $r_{\rm exp} \approx 10$~pc and follow a
de-Vaucouleur surface density profile with line-of-sight velocity
dispersions about 20~km/s.  A detailed description and analysis
of our merger-objects is subject of Fellhauer (2000) and a
follow-up paper.  Here we merely briefly note that our merger
objects fall in the region between dwarf galaxies and globular
clusters in the central surface brightness -- absolute magnitude
diagram (Fig.~3 in Ferguson \& Binggeli 1994).  Such objects
evolve into dSph-like systems in a periodic tidal field (Fig.~13
in Kroupa 1997; Fig.~2 in Kroupa 1998b).  More massive and
extended merger-objects can, of course, be obtained by suitable
choices of $R_{\rm pl}^{\rm sc}$ and $M_{\rm sc}$, but we defer a
more detailed discussion of this to the future.   

\section{Conclusions}
\label{sec:conclusions}

We have performed a set of self-consistent dynamical models of
clusters of twenty gas-free star clusters, as they have been
recently observed in the Antennae galaxies.  For all our models,
a central merger object formed out of a large fraction of the
single clusters.  As long as the super-cluster is smaller than
its tidal radius, almost all clusters merge. 

For our choice of parameters (5~kpc $\leq D \leq$ 100~kpc and
300~pc $\leq R_{\rm cut}^{\rm sc} \leq$ 1.8~kpc) the merger
process does not depend significantly on the two main
dimensionless parameters of the problem, the tidal field strength 
relative to the super-cluster concentration ($\beta$) and the
relative concentration of the super-clusters and the individual
clusters ($\alpha$).  The cluster merging process cannot be
modelled by a sequence of two-body merger events, but is a
``collective'' interaction, where the first passage
(``encounter'') of a cluster through the dense central region of
the super-cluster leads to the assimilation into a growing merger 
object.  The timescale of the merging process is the same
(measured in internal crossing times of the super-cluster $T_{\rm
  cr}^{\rm sc}$) for all models.  Measuring time in years shows
that high $\alpha$ calculations form the merger object within
very short times ($\approx 50$~Myr), while for extended clusters
(low $\alpha$) this process can take up to 1~Gyr or even longer.  

While for very strong tidal fields the tidal mass loss
as e.g. discussed by Baumgardt (1998)  dominates and merging
processes are suppressed, and for the limit where the individual
clusters approach point masses the merging is suppressed as well,
our parameter range, which is consistent with the observations,
always allows for the quick merger scenario on a few crossing
time scales.  Tidal mass loss is only a secondary effect for part
of our parameter space (stronger tidal field) due to a slight
reduction of the number of clusters available for merging.  
The number of merged clusters $n_{\rm m}$ decreases linearly with
$\beta$ if $\beta$ becomes larger than 1.0. 

We do not claim that our process of forming dwarf galaxies is the 
only possible way how these objects form, but at least it is one
possible way to explain the existence of low-mass dwarf galaxies
in the vicinity of large ``normal'' galaxies like our Milky Way.
In addition, some ``side-effects'' can be explained by our
models.  It takes very long until all clusters finally end up in
the merger object for low values of $\alpha$.  Such systems could
account for a high specific globular cluster frequency found in
dwarf spheroidal galaxies.  Our models develop tidal tails which
spread along the orbit of the dwarf galaxy.  Surviving escaped
star clusters are also found on the orbit.  This is similar to
what is observed for the Sagittarius dwarf spheroidal galaxy. 

\acknowledgements
Part of this project (parallelisation of the code, use of
high-performance computing power) was carried out at the
Edinburgh Parallel Computing Centre (EPCC) by the
TRACS-programme.  The TRACS-program\-me is a scheme of the
European Community (Access to Research Infrastructure action 
of the Improving Human Potential Programme; contract No 
HPRI-1999-CT-00026) which allows young Phd- or post-doc scientist
to learn parallel computing, offering computing time and support
for their projects.  MF thanks the staff of EPCC for their
support.  We also acknowledge the insightful discussions with
Prof.\ D.C. Heggie of the Department of Mathematics \& Statistics
at the University of Edinburgh.

$^{1}$: ASP Conf.\ Ser.\ 211: Proceedings of the workshop
"Massive Stellar Clusters" held in Strasbourg Nov.\ 1999; eds.\
A. Lancon, C.M. Boily
\end{article}
\end{document}